\documentclass[prc,amsmath,twocolumn,showkeys,showpacs,superscriptaddress]{revtex4}
\usepackage{graphicx,color} \usepackage{amssymb}
\usepackage{enumerate} \usepackage{verbatim} \usepackage{natbib}

\begin{document}
  \newcommand {\nc} {\newcommand} \nc {\Sec} [1] {Sec.~\ref{#1}} \nc
              {\IR} [1] {\textcolor{red}{#1}} 

\title{Role of core excitation in (d,p) transfer reactions}

\author{A. Deltuva} \affiliation{Institute of Theoretical Physics and
  Astronomy, Vilnius University, Saul\.etekio al. 3, LT-10222 Vilnius,
  Lithuania} \author{A.~Ross} \affiliation{National Superconducting
  Cyclotron Laboratory, Michigan State University, East Lansing, MI
  48824, USA} \affiliation{Department of Physics and Astronomy,
  Michigan State University, East Lansing, MI 48824-1321}
\author{E. Norvai\v{s}as} \affiliation{Institute of Theoretical
  Physics and Astronomy, Vilnius University, Saul\.etekio al. 3,
  LT-10222 Vilnius, Lithuania} \author{F.~M.~Nunes}
\affiliation{National Superconducting Cyclotron Laboratory, Michigan
  State University, East Lansing, MI 48824, USA}
\affiliation{Department of Physics and Astronomy, Michigan State
  University, East Lansing, MI 48824-1321}

\date{\today}


\begin{abstract}
\begin{description}

\item[Background:]  Recent work  found that core excitation can be
  important in extracting structure information from (d,p) reactions. 
\item[Purpose:]  Our objective is to systematically explore the role
  of core excitation in (d,p) reactions, and understand the origin of
  the dynamical effects.
\item[Method:] Based on the particle-rotor model of  $n+^{10}$Be, we
  generate a number of models with a range of separation energies
  ($S_n=0.1-5.0$ MeV), while maintaining a significant core excited
  component. We then apply the latest extension of the momentum-space based
  Faddeev method, including dynamical core excitation in the reaction
  mechanism to all orders, to the $^{10}$Be(d,p)$^{11}$Be like
  reactions, and study the excitation effects for beam energies from
  $E_d=15-90$ MeV. 
\item[Results:]  We study the resulting angular distributions and the
  differences between the spectroscopic factor that would be extracted
  from the cross sections, when including dynamical core excitation in
  the reaction, to that of the original structure model. We also
  explore how different partial waves affect the final cross section.
\item[Conclusions:] Our results show a strong beam energy dependence
  of the extracted spectroscopic factors that become smaller for
  intermediate beam energies. This dependence increases for loosely
  bound systems.
\end{description}
\end{abstract}

\pacs{21.10.Jx, 24.10.Ht, 25.40.Cm, 25.45.Hi}

\keywords{(d,p) reactions,  transfer reactions, core excitation}

\maketitle

\section{Introduction}

Throughout the history of nuclear physics, transfer reactions have
been important probes for nuclear structure and nuclear
astrophysics. Within this broad class, A(d,p)B reactions play a
prominent role due to the low Coulomb barrier and the well controlled
description of the deuteron. In the last two decades, (d,p) reactions
in inverse kinematics have been used to study properties of rare
isotopes. One example is the study of the one-neutron halo nucleus
$^{11}$Be; e.g. \cite{fortier1999,winfield2001, schmitt_prl2012,
  schmitt_prc2013}.

It is often the case that reaction theories freeze the degrees of
freedom in the core, and consider only the static effects of core
excitation by comparing the experimental cross sections with the
reaction theory predictions assuming a single particle structure
(e.g. \cite{schmitt_prc2013}). This is done mostly for simplicity
because the inclusion of core excitation  represents a  large increase
in complexity and sometimes poses conceptual issues in the
interpretation of data \cite{Lovell_jpg2015}. One may expect core
excitation to come into play at some point, but it is unclear under what
conditions these effects would be stronger. One may argue that at very
high energies, the collision time is too short for dynamical effects,
and that at very low energies, near and below the Coulomb barrier,
there is less probability for the core to be excited. What happens in
between these extreme limits is unclear. This is the topic we explore
in the present paper.

Core excitation effects have been considered in detail in the context
of the breakup of loosely bound two-body-like projectiles. The
three-body continuum discretized coupled channel method has been
extended to include core excitation to all orders and a variety of
applications
\cite{summers2006R,summers2006,summers2006b,summers2007,summers2008,moro2012,diego2014}
have all demonstrated that when including core excitation dynamically
in the reaction, breakup observables are significantly modified. In
some cases, the inclusion of core excitation helped describe specific
features of the data or even modified the physical interpretation
(e.g. \cite{diego2014}).  


Recently, the  full Faddeev formalism of \cite{Deltuva_prc2009} was
extended to include core excitation \cite{deltuva2013,deltuva2015}.
In \cite{deltuva2015}, the method is applied to
$^{10}$Be(d,p)$^{11}$Be and it is shown that dynamical core excitation
can be very important for the reaction populating the ground state of
$^{11}$Be, while less important for that populating the first excited
state.  These results called for further study.

In this work, we will use the method developed in \cite{deltuva2015}
and investigate the causes for the strong dynamical effects found in
that work. We are particularly interested in exploring the dependence
on the neutron separation energy and whether this is a phenomenon
unique to halo nuclei. Although the purpose of our work is not to
describe data, by  isolating the specific features that induce the
large core excitation effects, our work will help identify those
experiments for which a more computationally intensive analysis, fully
including core excitation in the reaction mechanism, may  be needed.

The paper is organized as follows. In Section II we provide a brief
description of the theory and the inputs required. Section III
includes all our results, from the $n+^{10}$Be toy models generated
(Section IIIa), to  angular distributions for elastic, inelastic and
transfer cross sections (Section IIIb) and the resulting extracted
structure information (Section IIIc). Finally, in Section IV we
present our summary and conclusions.


\section{Brief description of theory and inputs used}
\label{theory}

The Faddeev formalism for the description of three-body nuclear
reactions including core excitation and its numerical implementation
is taken over from Ref.~\cite{deltuva2015}.  It is based on the
integral form of the scattering theory as given by  the
Alt-Grassberger-Sandhas (AGS) equations \cite{alt:67a} for three-body
transition operators, extended for a multicomponent system, i.e.,
\begin{equation}  \label{eq:Uba}
U_{\beta \alpha}^{ji}  = \bar{\delta}_{\beta\alpha} \, G^{-1}_{0}  +
\sum_{\gamma=1}^3 \sum_{k=g,x}   \bar{\delta}_{\beta \gamma} \, T_{\gamma}^{jk}  \,
G_{0} U_{\gamma \alpha}^{ki}\;.
\end{equation}
The subscripts $\alpha, \beta, \gamma$ label the spectator particles
(interacting pairs in the odd-man-out notation), while 
the superscripts $i,j,k$ label the components of the operators 
coupling different  states of the core. Furthermore,
$\bar{\delta}_{\beta\alpha} = 1 - \delta_{\beta\alpha}$,
 $G_0 = (E+i0-H_0)^{-1}$ is the free resolvent at the reaction energy $E$, and
\begin{equation}  \label{eq:Tg}
T_{\gamma}^{ji} =  v_{\gamma}^{ji} +\sum_{k=g,x} 
v_{\gamma}^{jk} G_0 T_{\gamma}^{ki}
\end{equation}
 are  two-body
transition operators. The problem is formulated in an extended Hilbert
space with two sectors corresponding to ground ($g$) and excited ($x$) states of
the core.  These two sectors are coupled by  pairwise potentials
$v_{\gamma}^{ji}$, while the extended free Hamiltonian $H_0$ is diagonal in
the two sectors. Note that the kinetic energy operator also contains
the internal  core Hamiltonian. As a consequence, two- and three-body
transition operators couple the two sectors as well.  The amplitudes
for  three-body reactions are given by the on-shell matrix  elements
of $U_{\beta \alpha}^{ji}$ calculated between initial and final channel
states $|\Phi_\alpha^i \rangle$.
Denoting the core as particle 1 and proton
as particle 2, amplitudes for elastic deuteron scattering are given by
$\langle \Phi_1^g|U_{11}^{gg}|\Phi_1^g \rangle$,
for inelastic deuteron scattering by 
$\langle \Phi_1^x|U_{11}^{xg}|\Phi_1^g \rangle$, and for the
transfer reaction by 
$\langle \Phi_2^g|U_{21}^{gg}|\Phi_1^g \rangle + 
\langle \Phi_2^x|U_{21}^{xg}|\Phi_1^g \rangle$, since for the latter 
the final channel has two components.

Calculations are performed in momentum-space partial-wave
representation.  The proton-core Coulomb interaction is included via
the screening and renormalization method \cite{deltuva:05a}.  Except
for the partial waves with ${}^{11}$Be bound states,  described in
details in the next section, the pair interactions $v_{\gamma}^{ji}$ are
chosen following the strategy of  Ref.~\cite{deltuva2015}. A realistic
CD Bonn potential \cite{machleidt:01a} is used for the $np$ pair,
acting in partial waves with pair orbital angular momentum $L \leq 3$.
Nucleon-core potentials are based on the Chapel Hill 89 (CH89)
parametrization \cite{ch89}, but are deformed  with 
quadrupole deformation parameter $\beta_2 = 0.67$  and
deformation length  $\delta_2 = 1.664$ fm.
As in Ref.~\cite{deltuva2015}, a subtraction technique is used to preserve the 
 elastic nucleon-core cross section.
 The proton-core (neutron-core) interaction is included
in  partial waves with pair orbital angular momentum $L \leq 10$ 
($L \leq 5$).  The total three-body  angular momentum is limited to 
$J\leq 25$ which is sufficient for the convergence of elastic, inelastic
and transfer observables.

\section{Results}
\label{results}

\subsection{Models for ${}^{11}$Be}

The three-body Faddeev method of \cite{deltuva2015} assumes the final
nucleus can be represented within the particle-rotor model
\cite{nunes1996,nunes1996b,fthesis}. In this way, the ground state of
a nucleus like $^{11}$Be, would contain not only the $s_{1/2}$
components coupled to the $^{10}$Be ground state, but also $d_{3/2}$
and $d_{5/2}$ components coupled to the $^{10}$Be $2^+$ first excited
state (with excitation energy of $E_x=3.368$ MeV), resulting from the
quadrupole deformation of the core $^{10}$Be.   It is widely accepted that,
indeed, the ground state of $^{11}$Be contains a $\approx 20$\% core
excited d-wave admixture in the wavefunction.

To explore core excitation effects and its dependence on the
neutron separation energy, we needed to generate a variety of
$n+^{10}$Be models. Starting from the model developed in
\cite{nunes1996b} one can produce a variety of models with different
separation energies just by changing the depth of the central
interaction while keeping the geometry fixed. It is important to
ensure that all models produce a similar significant admixture as the
original $^{11}$Be model of \cite{nunes1996b}. If the $n+^{10}$Be
system exhibits a larger core excited component one might induce
larger core excitation effects in the reaction as a consequence of the
structure, rather than the reaction mechanism.  While most models
include a core with the physical excitation energy of the $2^+$ state
in $^{10}$Be, we also explored the effect of a small excitation
energy $E_x=0.5$ MeV. There are a couple of $n+^{10}$Be models in
Table I developed for this purpose.

The resulting potentials are presented in Table I. We include the
central depth $V_{ws}$, the spin-orbit depth $V_{so}$, the
deformation length $\delta_2=\beta_2 R_{ws}$ and the core $2^+$ excitation energy $E_x$. The geometry of the
central and spin-orbit force are kept as in \cite{nunes1996b}, namely
$R_{ws} = 2.483$ fm, $a_{ws} = 0.65$ fm. The last two columns in Table
I correspond to the percent probability associated with the dominant partial
waves included in the $n+^{10}$Be model space (also referred to as the
theoretical spectroscopic factors). These results were obtained with
{\sc Efaddy}~\cite{efadd} by solving the coupled channel
equation \cite{nunes1996}, but were verified also by momentum-space calculations.

\begin{table}[b]
\centering
\begin{tabular}{|r | r r r r | r r|}
\hline $ S_n $(MeV) & $V_{ws}$(MeV)    & $V_{so}$ (MeV) &$\delta
\;$(fm) &E$_x$(MeV)&  $S^{th}_{s1/2}$ &    $S^{th}_{d5/2}$\\ \hline
0.1 & -51.924 & -8.5 & 1.664 & 3.368 & 94.2 &  5.4\\
\textbf{0.5}& \textbf{-54.45}
&\textbf{-8.5}&\textbf{1.664}&\textbf{3.368}&\textbf{85.4}&\textbf{12.5}\\ 
0.5&-52.988&-1.0&1.664&0.500&79.2&13.3\\ 1.0&
-56.475& -8.5&   1.664&3.368&78.7&18.4\\
5.0& -67.059& -8.5&
1.664&3.368&57.7&37.4\\ 5.0&-65.670&-1.0&1.664&0.500&54.5&29.7\\
\hline
\end{tabular}
\caption{Parameters for the n$+^{10}$Be system taking into account
the  $^{10}$Be($2^+$) core excitation, and the resulting spectroscopic
  factors, as a function of the neutron separation energy $S_n$.}
\end{table}

For comparing the core excitation results with those obtained under
the assumption of a single-particle structure, we also produce the
corresponding $n+^{10}$Be single-particle potentials. Here we have
imposed volume conservation as discussed in \cite{nunes1996}.  The
central depths of the resulting potentials are summarized in Table II. 

\begin{table}[b]
\centering
\begin{tabular}{|r | r |}
\hline $ S_n $(MeV) & $V_{ws}$(MeV) \\ \hline 0.1& -57.319  \\ 0.5&
-61.243 \\ 1.0& -64.337 \\ 5.0& -79.378 
\\ \hline
\end{tabular}
\caption{Single particle parameters for n$+^{10}$Be system, as a
  function of the neutron separation energy $S_n$. The depth of the
  spin-orbit force is the same for all these models $V_{so}=8.5$ MeV.}
\end{table}

\subsection{Predicted cross sections}

\begin{figure}[t]
\center \includegraphics[scale=0.27]{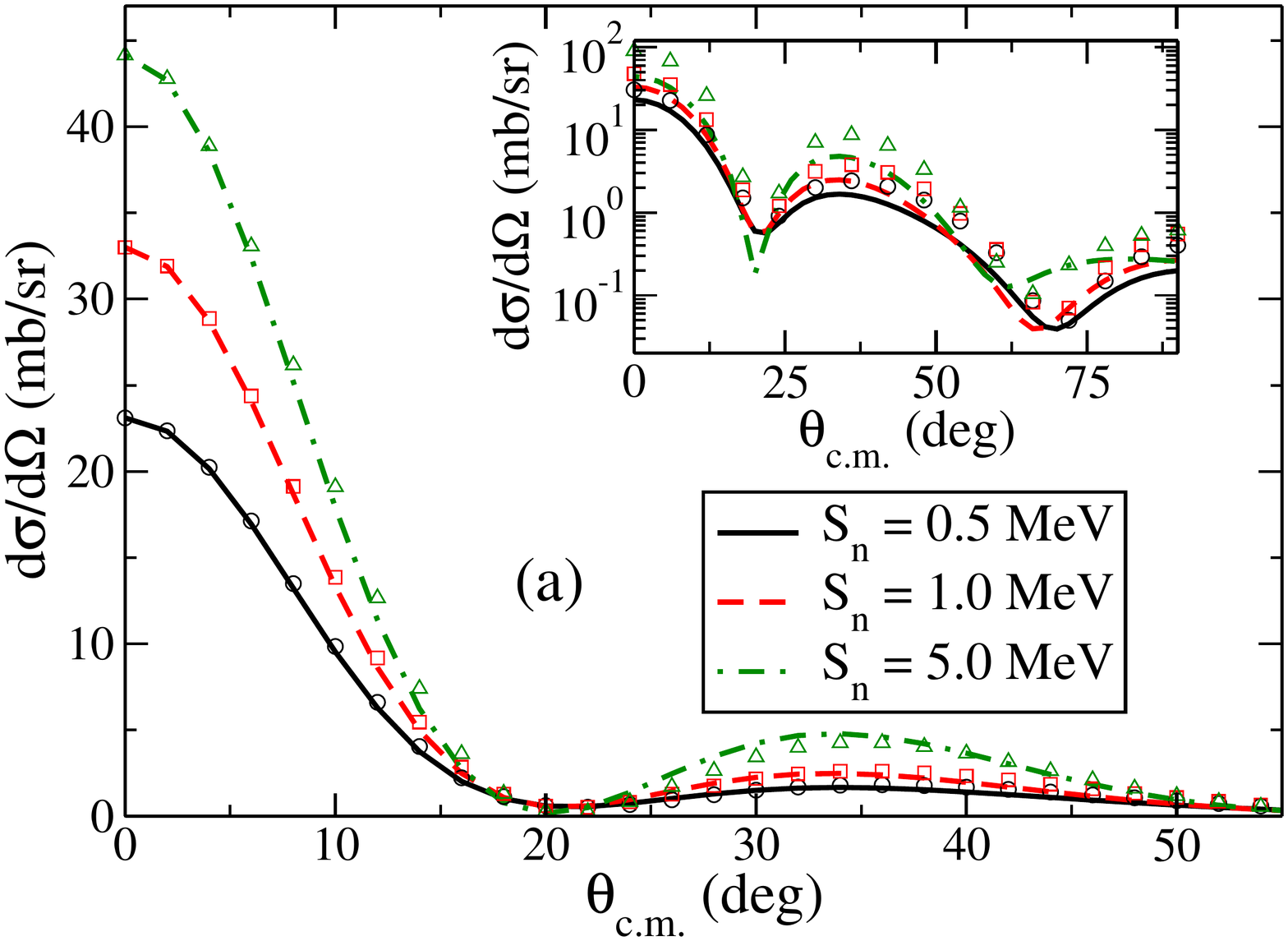}
\includegraphics[scale=0.27]{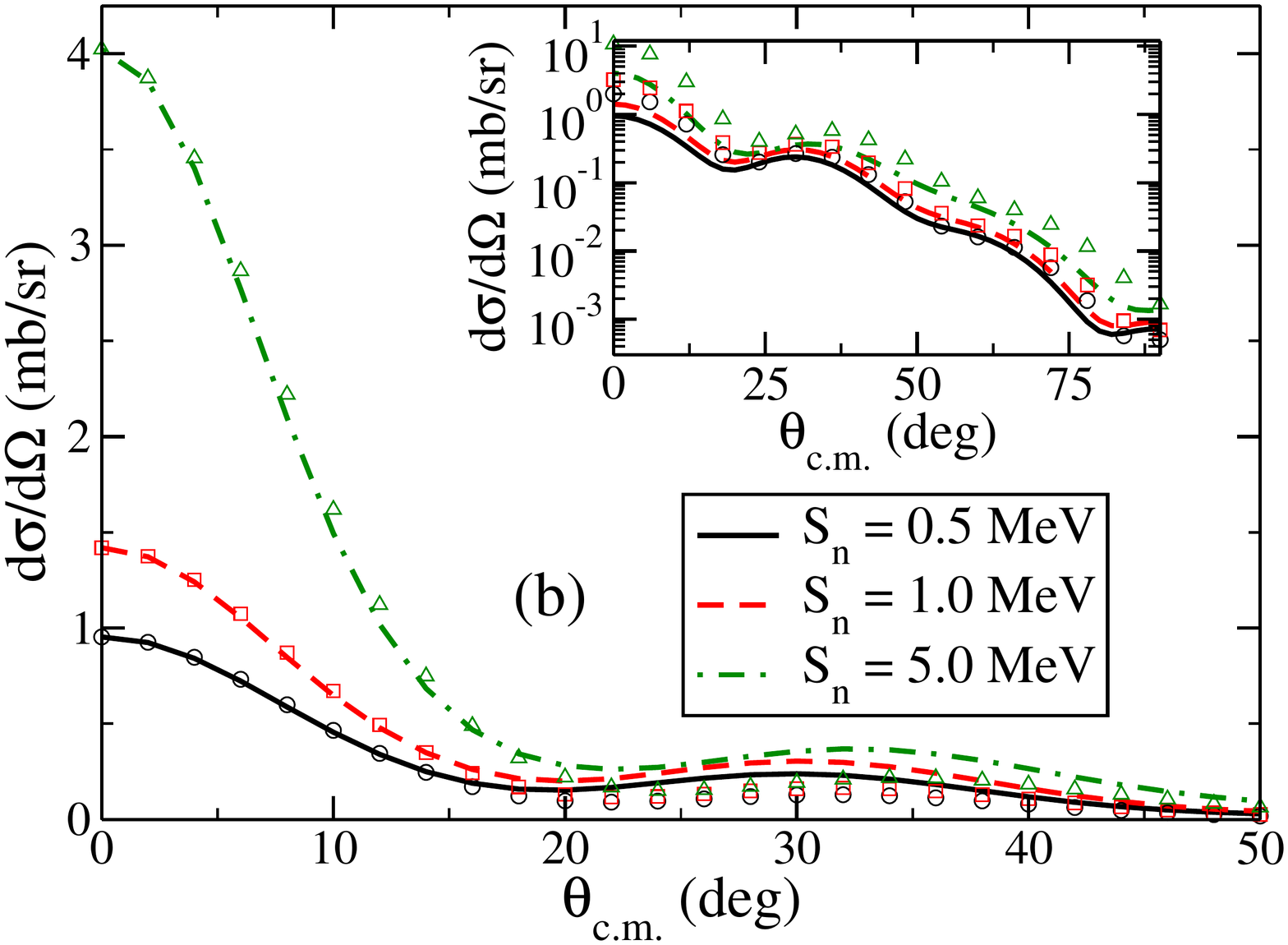}
\caption{(Color online) Angular distributions for
  $^{10}$Be(d,p)$^{11}$Be(g.s.) at 20 MeV (a) and 80 MeV  (b) for
  various separation energies of the final nucleus.}
\label{transfer-xs}
\end{figure}

The magnitude of the transfer cross sections depends strongly on the
beam energy and their angular distribution provides information on the internal orbital
angular momentum of the final nucleus. For a given beam energy, the
transfer cross section is largest when there is optimum Q-value
matching. In Figs.\ref{transfer-xs}(a) and (b) we show the angular
distributions for $^{10}$Be(d,p)$^{11}$Be for deuteron energies of 20
MeV and 80 MeV, respectively, produced with the full Faddeev method
including core excitation: the solid black line represents the
prediction for the model with the realistic separation energy, while
the red-dashed and the green-dot-dashed correspond to a separation
energy of $S_n=1$ MeV and $S_n=5$ MeV, respectively. The distributions
are all forward peaked because they correspond to $L=0$
transitions. The insets in Figs.\ref{transfer-xs}(a) and (b)
correspond to the same data but instead in log plot.  In addition to
the predictions including fully dynamical core excitation, we include,
in symbols, the predictions obtained neglecting core excitation
(circles for $S_n=0.5$ MeV, squares for $S_n=1$ MeV and triangles for
$S_n=5$ MeV).

Focusing now on the log plots of Figs.\ref{transfer-xs} (a) and (b),
the main effect of core excitation is to reduce the cross
section. This is reflected in the fact that the normalizations needed to match the single-particle
predictions with the core excited predictions are smaller than unity. For example, for $S_n=0.5$
MeV, the normalization needed for the single-particle cross sections for $E_d=20$ MeV is $SF=0.76$, while for
$E_d=80$ MeV it is $SF=0.48$. We will come back to this discussion in
Section \ref{sec:rx}.

In the linear plots only, the single-particle predictions have been
normalized by an arbitrary factor to match the full core excited
predictions at zero degrees, to make the comparison of the shapes of
the distributions easier.  At low beam energies, minor changes in the
shape of the angular distribution are seen. However at the higher
energies there is a significant change in the shape for larger angles.

\begin{figure}[t]
\center \includegraphics[scale=0.27]{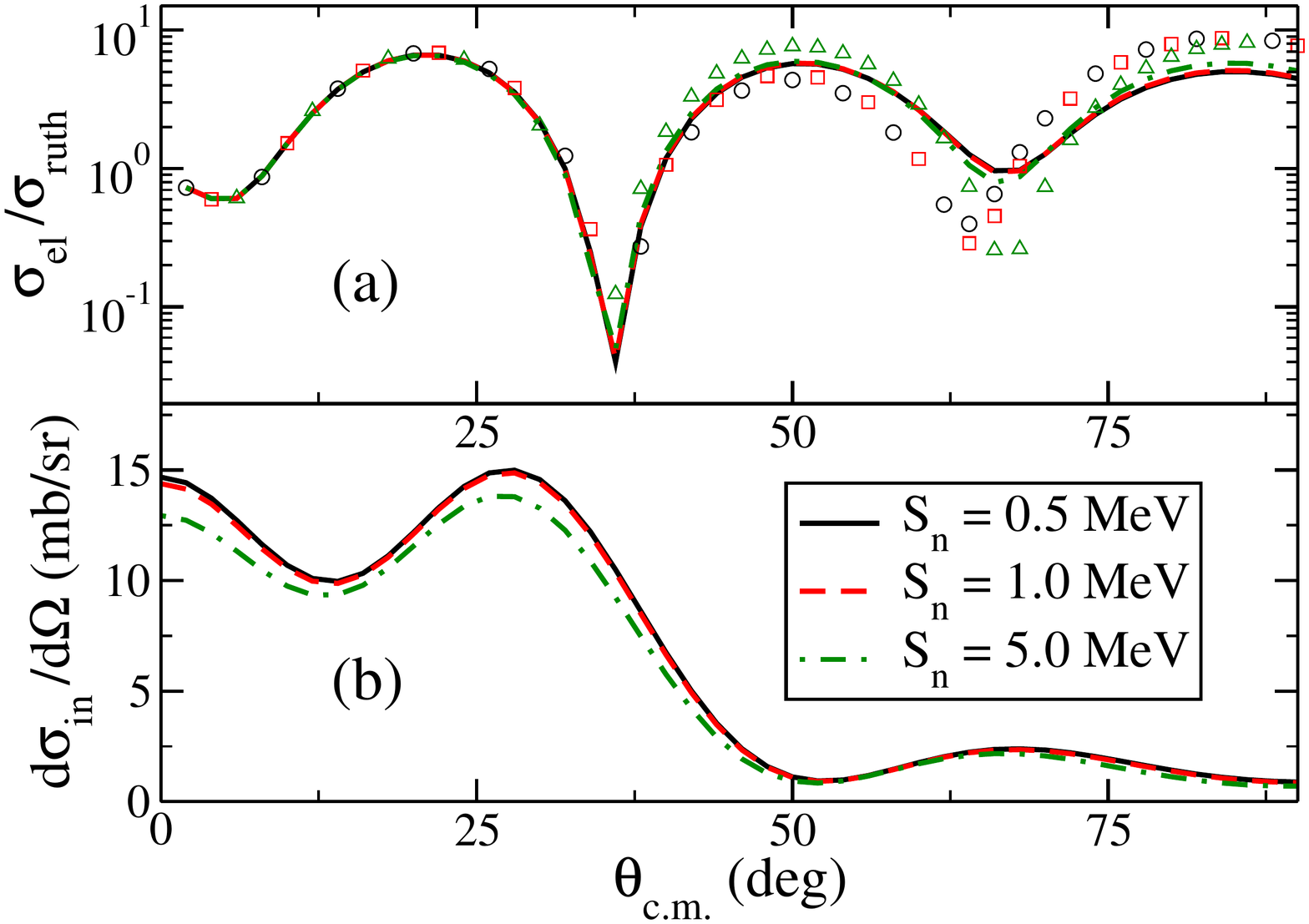}
\includegraphics[scale=0.27]{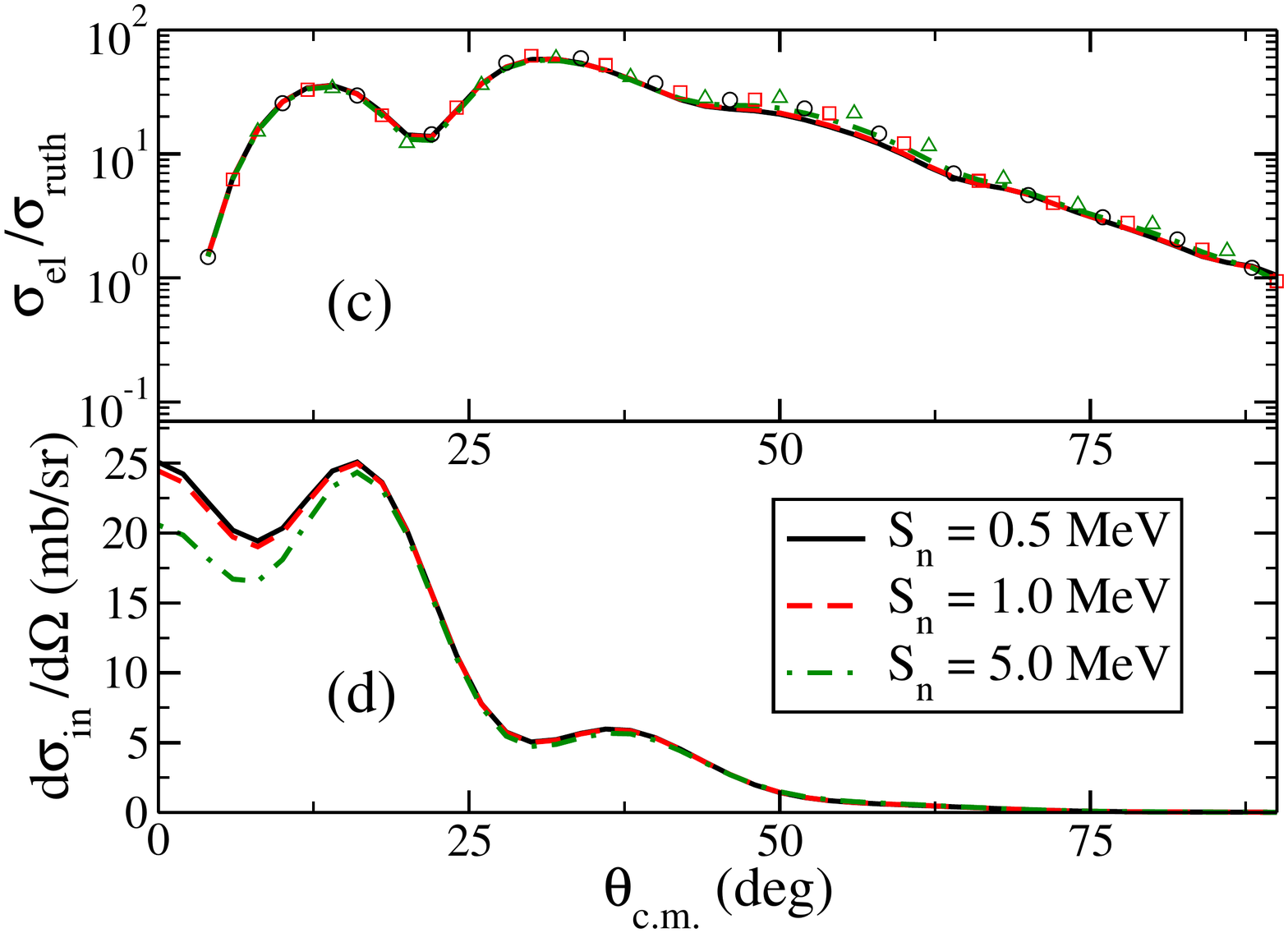}
\caption{(Color online) Angular distributions for
 $^{10}$Be(d,d)$^{10}$Be (ratio to Rutherford) and
  $^{10}$Be(d,d')$^{10}$Be$2^+$ at 20 MeV (a) and (b) and 80 MeV (c)
  and (d): Faddeev predictions are obtained for various separation energies of the final nucleus in the (d,p) transfer channel.}
\label{elastic-xs}
\end{figure}

The nucleon-target interactions determine the details of the elastic
and inelastic distributions.  In the full Faddeev calculations, these
predictions are produced consistently with the transfer
predictions. For completeness, we show in Fig.\ref{elastic-xs} the elastic 
and inelastic cross sections as functions of
scattering angle for $E_d=20$ MeV and $E_d=80$ MeV.  The full Faddeev
predictions with core excitation are shown by the lines:  solid black
line for $S_n=0.5$ MeV, dashed red line for $S_n=1$ MeV, and green
dot-dashed line for $S_n=5$ MeV.  The plots for the elastic
distributions also contain the single-particle predictions in symbols
(black circles for $S_n=0.5$ MeV, red squares for $S_n=1$ MeV, and
green triangles for $S_n=5$ MeV).  The elastic distribution is not
very sensitive to the separation energy of the system, but we see that
the inelastic cross section decreases with  increasing separation
energy, a consequence of the fact that the transition operator is
mostly sensitive to the surface of the optical potential (the operator
is roughly proportional to the derivative of the optical potential)
and therefore is enhanced when the composite nucleus has large tails
(small separation energies). Note that the single-particle model
predicts no inelastic cross sections. 
An experiment that measures all three channels (elastic,
inelastic and transfer) simultaneously will provide stringent
constraints to the reaction model. 

Of course, in addition, the Faddeev method with core excitation also
predicts elastic breakup cross sections (which leave $^{10}$Be in its
ground state) and inelastic breakup cross sections (which leave
$^{10}$Be in is $2^+$ excited state). However, it is far more
demanding  to obtain convergence for these observables and it is
beyond the scope of this work.

\subsection{Extracted structure information}
\label{sec:rx}

Often (d,p) transfer experiments are performed with the objective of
extracting a spectroscopic factor. This is done by taking the ratio of
the measured cross section at the peak of the angular distribution and
the corresponding theoretical prediction, assuming a pure
single-particle final state:
$S^{exp}=\frac{d\sigma^{exp}}{d\Omega}/\frac{d\sigma^{sp}}{d\Omega}$. Let
us consider specifically the realistic case of
$^{10}$Be(d,p)$^{11}$Be. We understand that the realistic overlap
function for the ground state of $^{11}$Be has in addition to the
neutron $s_{1/2}$ wave coupled to $^{10}$Be$(0^+)$, components where
the core is excited (in our model only $^{10}$Be($2^+$) is
considered). Due to the excitation energy of the core $E_x=3.368$ MeV,
the tails of the overlap function for those core-excited components
die off much faster than the $s_{1/2}$ component, which has an exponential
decay dominated by $S_n=0.5$ MeV. For that reason, it is this $s_{1/2}$ 
component that dominates the final cross section. If no dynamical core
excitation takes place during the reaction, then the ratio of cross
sections $S^{exp}$ corresponds exactly to $S^{th}_{s1/2}$, the
probability that the valence neutron is indeed in the $s_{1/2}$
single-particle orbital in the $^{11}$Be-like system. However,
dynamical effects in the reaction can change this value and produce
erroneous conclusions when extracting the spectroscopic factor from
transfer angular distributions. 
\begin{figure}[t]
\center \includegraphics[scale=0.27]{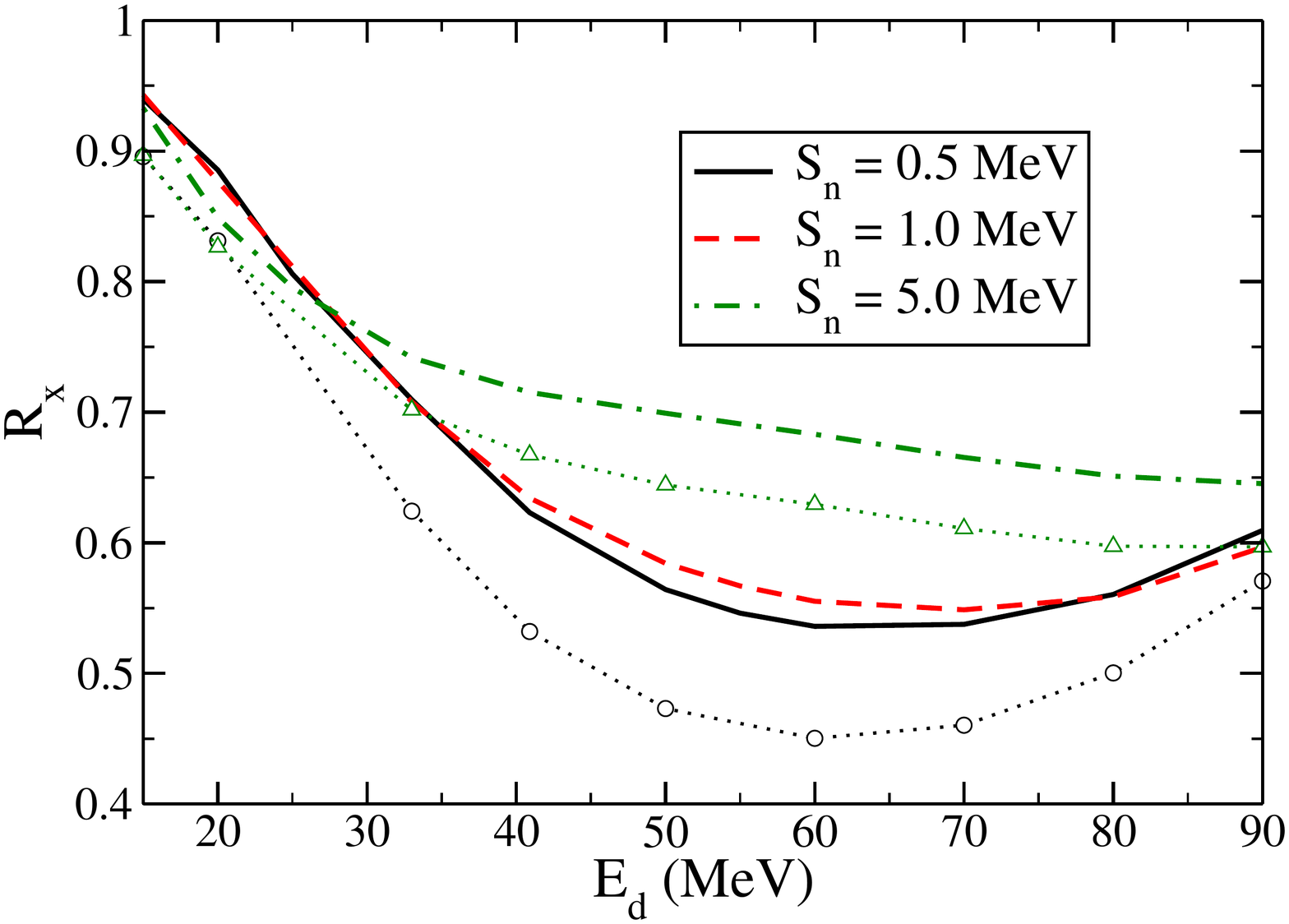}
\caption{(Color online) Spectroscopic factor ratio $R_x$ as a function
  of the beam energy extracted from $^{10}$Be(d,p)$^{11}$Be for
  various separation energies of the final nucleus (more details in
  the text).}
\label{rx}
\end{figure}

We test this idea using the full Faddeev predictions which include
core excitation to all orders. These predictions serve as our data
and, by comparing them with the single particle predictions, we can
extract $S^{Fadd}$ as a ratio of the Faddeev cross section when core
excitation is included and the single-particle Faddeev prediction, i.e.,
$S^{Fadd}=\frac{d\sigma^{cex}}{d\Omega}/\frac{d\sigma^{sp}}{d\Omega}$. We
then take the ratio of this spectroscopic factor $S^{Fadd}$ and  the
spectroscopic factor $S^{th}_{s1/2}$ introduced in our $^{11}$Be structure
model (Table I). This quantity is defined as $R_x=S^{Fadd}/S^{th}_{s1/2}$ and
is plotted in Fig. \ref{rx} for several separation energies as a
function of beam energy: $S_n=0.5$ MeV (solid black), $S_n=1$ MeV
(dashed red) and $S_n=5$ MeV (dot-dashed green). If there were no
dynamical excitations in the reaction, $R_x$ should be unity,
independent of the beam energy. We find that $R_x$ is not unity and
depends strongly on beam energy as was already indicated in
\cite{deltuva2015}. 

The coupling between components with different core states (in this
case a quadrupole term) is peaked at the surface; as mentioned before
the operator is roughly the derivative of the optical potential. For
cases in which the neutron separation energy is considerably smaller
than the excitation energy, one might naively expect two limiting
cases: at very small beam energies, dynamical effects should be small
because the Coulomb barrier keeps the deuteron far from the target not
allowing the nuclear quadrupole coupling to act, and at very high beam
energies, dynamical effects should also decrease because the timescale
for the reaction hinders multistep effects. 
It is for intermediate
beam energies that one can expect dynamical effects to take
place. This is exactly what is seen in Fig. \ref{rx}. Unfortunately
there are numerical difficulties in obtaining converged Faddeev
calculations for beam energies lower than $E_d=15$ MeV. The solid line
starts at $R_x=0.94$ for $E_d=15$ MeV, which is well above the Coulomb
barrier. It then decreases to a maximum effect around $E_d=60-70$ MeV
($R_x=0.54$), raising again for the higher beam energies. Extracting a
spectroscopic factor from data in the range $E_d=40-90$ MeV using the
single-particle predictions can lead to very large underprediction of
the spectroscopic factor.

Fig. \ref{rx} also shows that the effect of dynamical core excitation
is more pronounced for the more loosely bound systems.  This result
appears at first counter intuitive: if the transferred neutron moves
into a loosely bound orbital, it may not feel the effects of core
excitation as much. This is not the case: the larger the separation
energy, the smaller the strength of the coupling to excited states
(due to a weaker overlap of the $^{11}$Be-like components and the
transition operator). This manifests as a weaker dependence of $R_x$
as a function of beam energy. 
Although the dynamical effects described by the full Faddeev
equations are highly non-linear, we found that the dependence of $R_x$
on the separation energy $S_n$ is approximately linear except when the
separation energy approaches zero. The value of the separation energy
for this change in behavior depends on $E_d$ (for example, at $E_d=41$
MeV the linear behavior extends down to $S_n=0.3$ MeV).

Reducing the excitation energy of the core, increases the overlap of
the $^{11}$Be-like core-excited components and the transition
operator.  In Fig. \ref{rx} we also show the predictions when
$E_x=0.5$ MeV.  For both $S_n=0.5$ MeV (black circles) and $S_n=5$ MeV
(green triangles), the dynamical effects of core excitation are
enhanced when $E_x$ decreases, as demonstrated by the fact that the extracted ratios
$R_x(E_d)$ for $E_x=0.5$ MeV are below the $R_x(E_d)$ lines corresponding to
$E_x=3.368$ MeV.

\begin{figure}[t]
\center \includegraphics[scale=0.27]{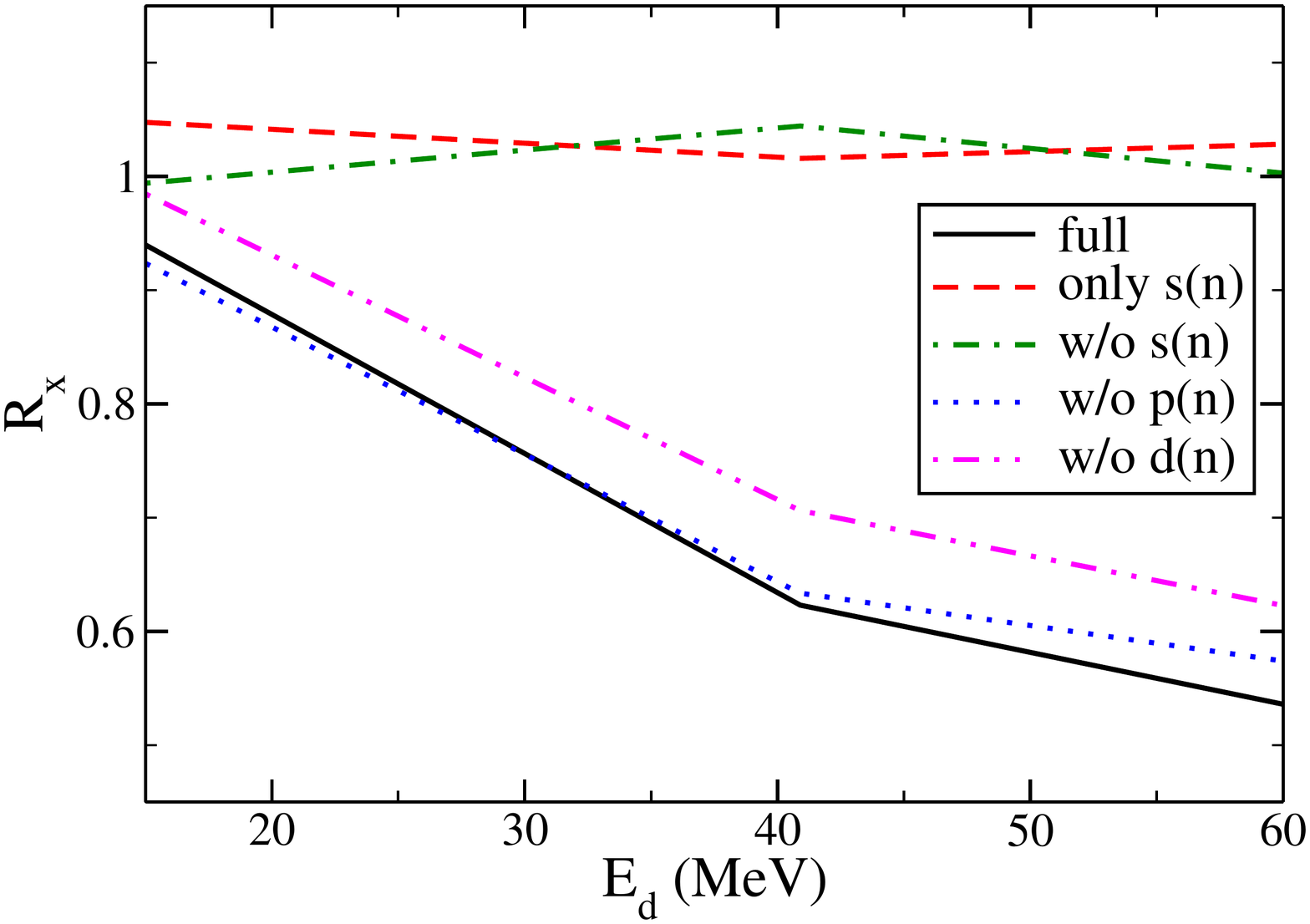}
\caption{(Color online) Spectroscopic factor ratio $R_x$ as a function
  of beam energy extracted from $^{10}$Be(d,p)$^{11}$Be, switching off
  various couplings (see text for more detail).}
\label{rx-spd}
\end{figure}

If we just include core excitation in the $^{11}$Be bound state, $R_x \approx 1$
 for all beam energies, which demonstrates that breakup is
critical to enable the dynamical effects we observe.  We then
investigate the different partial waves in the nucleon-target
subsystem that are responsible for the effect. In Fig. \ref{rx-spd} we
show the predictions for $R_x$ when only the s-wave is included in the
neutron-target relative motion (dashed red line) and compare it to the
results including all partial waves (solid black line). We see that
including only s-wave in the neutron-target continuum produces
virtually no effect. On the other hand, switching off the s-wave,
while keeping all other components in the calculation (dot-dashed
green line) also produces no effect. We also show the results obtained
when switching off the p-wave (dotted blue line) and the d-wave
(dot-dot-dashed purple line). It is the interplay of both s-waves and
d-waves that causes the large reduction observed for $R_x$.
We find that, for the proton-target interaction,  many partial waves are needed for convergence but 
there is no strong interference between them, as for the neutron-target case.

In analogy with the previously studied nucleon-deuteron scattering including dynamically the
$\Delta$-isobar excitation \cite{deltuva:03c}, 
the core excitation effect can be separated into contributions of two- and three-body nature. 
Inclusion of only the two-body part is possible within standard AGS equations with effective 
transition operators acting in a single sector (g) of the Hilbert space. The results of this
 calculation typically go in the opposite direction as compared to the full calculation,
thereby indicating that there is a strong competition between the contributions of two- and three-body 
nature, and the full core excitation effect is a result of a complicated interplay between them.
These findings are in qualitative agreement with those of
nucleon-deuteron scattering with the $\Delta$-isobar excitation \cite{deltuva:03c}.

Our NN interaction is the full CD-Bonn which contains the tensor force
and produces a deuteron bound state with the appropriate d-wave
component. Although we focused our study in the $^{11}$Be structure,
we also wanted to explore whether the NN tensor interaction could  be
contributing to the dynamical effects we observe.  We repeated the
calculations (including core excitation and assuming a single-particle
structure) using a simple Gaussian interaction for the NN force in the deuteron partial wave
(as in \cite{Upadhyay_prc2012}). The $R_x$ obtained in this way are
within $2$ \% of those obtained with the full CD-Bonn. We thus
conclude that the NN tensor force is not responsible for the dynamical
effect under study.


\section{Summary and Conclusions}
\label{conclusions}

Our main goal in this work is to systematically explore the role of
core excitation in (d,p) reactions, and understand the origin of the
dynamical effects.  We generate a number of two-body $n+^{10}$Be
models for a $^{11}$Be-like system, with a range of neutron separation
energies ($S_n = 0.1 - 5.0$ MeV), keeping a significant core excited
component. We then perform full Faddeev calculations including core
excitation to all orders to obtain elastic, inelastic and transfer
cross sections.  We study the effect of core excitation and extract
the spectroscopic factor that would be obtained taking the ratio of
the full calculation with that of the single-particle model, for a
range of beam energies. 

The spectroscopic factors obtained by taking this ratio do not agree
with the spectroscopic factor in the original model. For example, in
the case of the realistic $^{11}$Be, the spectroscopic factor obtained
is strongly dependent on beam energy, with a minimum of half its
original value at intermediate beam energies of around $60-70$
MeV. All this points towards the fact that dynamical core excitation
is indeed distorting the results and should be explicitly included in
the reaction mechanism for a reliable extraction of structure
information.  Increasing the neutron separation energy, reduces the
effect. 

In order to fully explore this dynamical effect we also perform
calculations when the excitation energy of the core is arbitrarily
reduced to $0.5$ MeV. This reduction increases the role of core
excitation, regardless of the separation energy of the system, or the
beam energy considered, since then the core excited components in the
$^{11}$Be-like system have an asymptotic fall-off comparable to the
component where the core is in its ground state, enhancing
core-excitation couplings.

Finally, we also explore the role of different partial waves in the
nucleon-target subsystem in the final transfer cross section.  We find
that interference effects between s-wave and d-waves in the
neutron-target continuum are essential to reproduce the full result.

This interesting phenomenon of dynamical core excitation is
sufficiently large that it merits experimental investigation. The
reaction $^{10}$Be(d,p)$^{11}$Be has been measured in the lower energy
regime \cite{schmitt_prc2013}. Unfortunately the magnitude of the
transfer cross section decreases significantly with beam
energy. However, a measurement in the intermediate energy range
$E_d=80$ MeV may still be feasible and would provide a crucial test on
the predictions of the reaction model, particularly if the various
reaction channels are measured simultaneously as in
\cite{schmitt_prc2013}.

{ 
A similar investigation of the role of core excitation in Eikonal models for nuclear knockout reactions and 
its dependence on beam energy may shed light on the reductions factor observed when extracting structure information from those measurements \cite{tostevin2014}. 
}


\begin{center}
\textbf{ACKNOWLEDGMENTS}
\end{center}

We are grateful to Charlotte Elster for  useful discussions. 
The work of A.D. and E.N. was supported  by Lietuvos Mokslo Taryba 
(Research Council of Lithuania) under contract No. MIP-094/2015.
 The work of A.R. and F.M.N.
 was supported by the National Science Foundation under Grants
No. PHY-1520929 and PHY-1403906 and the Department of Energy under
Contract No. DE-FG52- 08NA28552.



\end{document}